# New phenomena in CDW systems at small scales

S.V. Zaitsev-Zotov

*Kotel'nikov Institute of Radio Engineering and Electronics of Russian Academy of Sciences*

+7(495)6293394, serzz@cplire.ru

The first observation of finite-size effects was reported by Borodin *et al.* [1] more than 24 years ago. Since then tens of new finite-size effects were found in small samples of charge-density wave conductors. Study of finite-size effects changed many of our initial expectations and helped to resolve many problems: we know now that the phase-correlation lengths are in micrometer region, that pinning of the CDW is collective *etc*. Here we present a brief summary of the main achievements in this area.

Two characteristic lengths are relevant to the finite-size effects: the phase-correlation length, $L$ (also known as Larkin-Ovchinnikov or Fukuyama-Lee-Rice length), and the amplitude correlation length, $\xi$ ($\sim \hbar v_F/\Delta$). The typical values for $L$ are relatively large ($L_\parallel \sim 10$ μm, $L_\perp \sim 1$ μm for o-TaS$_3$, NbSe$_3$) and can be easily achieved experimentally. Finite-size effects are observed in samples having at least one dimension below the characteristic length.

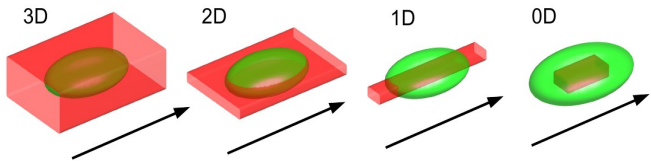

**Figure 1.** Dimensionality of finite-size effects for the most usual cases. A phase-correlation volume is shown in green, a sample is red. Arrows denote the chain direction.

Figure 1 shows a simple classification of dimensionality of the finite-size effects. Circles in figures 2 and 3 show the progress in reduction of samples cross-sectional areas and inter-contact distances (Refs. [1-11]) respectively since the first publication in this field. Surprisingly, the progress follows the Moor's law well-known for semiconductor industry (blue lines). We would like to note that there was no real progress in reduction of both sample sizes and contact separations for more than 5 years. Despite of this, a number of very interesting observations were reported recently. The list of finite-size effects presented below includes some recent illustrations and renewal of some statements of the finite-size effects. More detailed description of earlier results can be found in the review [12].

One of the most noticeable manifestation of the finite-size effect is growth of the threshold field for the onset of nonlinear conduction, $E_T$, as a function of transverse sizes of a crystal. The functional form depends on the pinning dimensionality as follows:

2D: $E_T \propto t^{-1}$, where t is the sample thickness [13,14];
1D: $E_T(s) \propto s^{-2/3}$, where $s$ is the sample cross-sectional area [14];
0D: $E_T(s) \propto v^{-1/2}$, where $v$ is the sample volume [2].

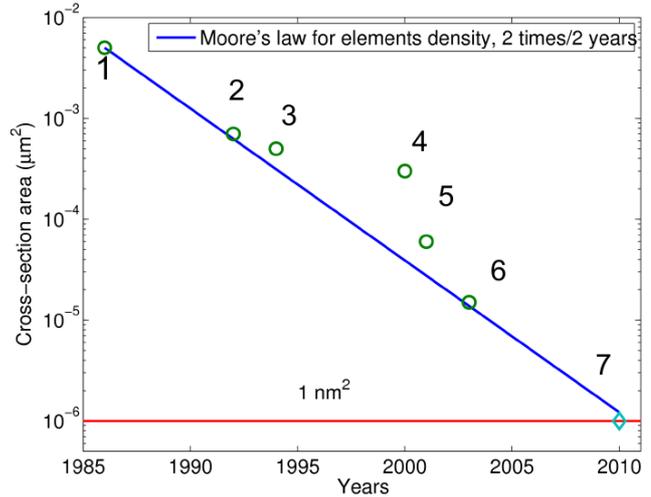

**Figure 2.** Progress in reduction of cross-sectional area of CDW crystals. The references are given by numbers.

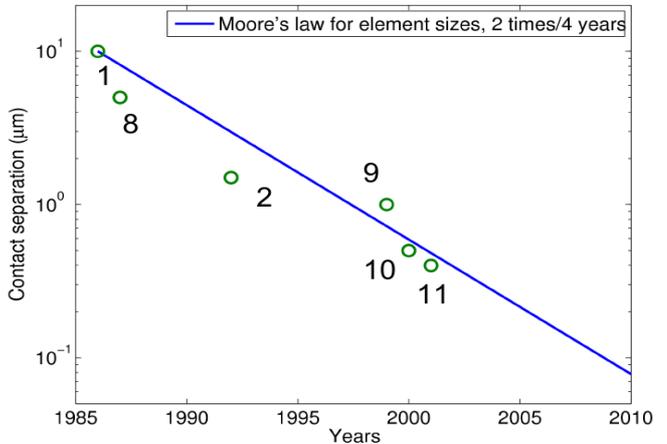

**Figure 3.** Progress in reduction of inter-contact distances (only working contacts are taken into account).

Fine details of CDW reordering, which reveal itself in jump-like resistance changes, can be observed in small crystals [1,3]. The detailed analysis of such jumps allows precise study of temperature variation of the charge-density wave vector and provides an estimate of mobility of the current carriers [15].

Low-temperature collective transport in thin CDW conductors is also specific: temperature-independent I-V curves observed at temperatures below 10 K follow unusual law, $I \propto \exp(-[V_0/V]^D)$ [16] with crystal-shape dependent exponent D = 1, 2 [17] and can be described in terms of various macroscopic quantum phenomena [18-21].

Fluctuations are bigger in thin crystals. They lead to smearing out of the Peierls transition [7,8,13], appearance of spontaneous fluctuations of resistance [22]. Surprisingly, smoothing of the threshold behavior below the Peierls transition temperature $T_P$ [13] is accompanied by appearance of the threshold nonlinear conduction above $T_P$ [23]. As thin crystals have very good thermal contact with a substrate, they allow to observe transport phenomena at very big current density, corresponding to the gigahertz region [24].

Further reduction of sample thickness below 0.1 μm leads to qualitative changes of physical properties. The most bright manifestation is observed in NbSe$_3$ where thin crystals lose their metallic properties and turn to dielectric behavior $R \propto T^{-\alpha}$ accompanied by development of nonlinear I-V curves $I \propto V^\beta$ [5,6,25]. Such dependencies are expected for various kinds of 1D systems including Luttinger liquid stabilized by impurities [26]. A contribution from the surface CDW [7] may also take place.

If a thin crystal is also short, we come to 0D pinning case where $E_T$ exhibits mesoscopic variations with temperature [2]. Shortening of distance between current terminals leads also to decrease of the phase-slip voltage [10], whereas shortening of distance between voltage probes reveals negative resistance regions [11].

Some surface phenomena can be observed in thin crystals only and can be considered as a sort of finite-size effect. In particular, CDW transport in thin crystals is sensitive to the transverse electric field [3]. Surface of CDW conductors is sensitive to illumination which substantially changes both the collective transport [27,28] and linear transport [28-30]. The latter allows to study the energy structure of the Peierls conductors [30] and even distinguish between collective and single-particle low-temperature transport [29]. And, of course, surface of CDW conductors is a separate object having other physical properties than CDW in a bulk [7]. Successful investigation of these properties means that we are already on the amplitude-correlation length scale (diamond in Fig. 2).

If we are talking about nanoscale phenomena, we should mention various effects related to nanoscale structures in bulk CDW crystals. CDW crystals with columnar defects demonstrate interference phenomena [31,32]. I-V curves of point contacts [33], inter-layer tunneling [34,35], nanoconstrictions and break-junctions [36] have features of the energy structure.

As it is clear from above, new technologies and new methods provided us with a lot of new physics on small scales. We believe that this is the time now when the finite-size effect can be used as a tool for further investigation of very reach properties of CDW conductors.